\documentclass{elsart}
\usepackage[dvips]{graphicx}
\usepackage{amsmath}
\usepackage{amssymb}
\usepackage{epsfig}
\usepackage{psfrag}
\usepackage{color}
\usepackage{graphicx}
\usepackage{color,ulem}

\begin{document}
\begin{frontmatter}

\title{Finite Size Effects for the Ising Model on
    Random Graphs with Varying Dilution}

\author{Julien Barr\'e$^{1}$},
\author{Antonia Ciani$^{2,4}$},
\author{Duccio Fanelli$^{3,4}$},
\author{Franco Bagnoli$^{3,4}$},
\author{Stefano Ruffo$^{3,4}$}

\address{1. Laboratoire J.A. Dieudonn\'e, UMR CNRS 6621, Universit\'e de Nice-Sophia Antipolis, Parc Valrose
06108 Nice, France}
\address{2. Dipartimento di Fisica, Universit{\`a} di Firenze, and INFN, Via Sansone 1, 50019 Sesto F.no (Firenze), Italy}
\address{3. Dipartimento di Energetica and CSDC, Universit\`a di Firenze, and INFN, via S. Marta, 3, 50139
Firenze, Italy}
\address{4. CSDC,  Centro interdipartimentale per lo Studio delle Dinamiche Complesse, 
Universit{\`a} di Firenze, Via Sansone 1, 50019 Sesto F.no (Firenze), Italy} 

\date{\today}

\thanks[julien]{E-mail: jbarre@math.unice.fr}
\thanks[antonia]{E-mail: antoniaciani@gmail.com}
\thanks[duccio]{E-mail: duccio.fanelli@unifi.it}
\thanks[franco]{E-mail: franco.bagnoli@unifi.it}
\thanks[stefano]{E-mail: stefano.ruffo@unifi.it}

\begin{abstract}
We investigate the finite size corrections to the 
equilibrium magnetization of an Ising model on a random graph with $N$ nodes and
$N^{\gamma}$ edges, with $1 < \gamma \leq 2$. By conveniently
rescaling the coupling constant, the free energy is made extensive. As
expected, the system displays a phase transition of the mean-field
type for all the considered values of $\gamma$ at the transition
temperature of the fully connected Curie-Weiss model. Finite size
corrections are investigated for different values of the
parameter $\gamma$, using two different approaches:
a replica-based finite $N$ expansion, and a cavity method.
Numerical simulations are compared with theoretical predictions. The
cavity based analysis is shown to agree better with numerics.
\end{abstract}

\end{frontmatter}

\section{Introduction}
Complex networks of interacting elements are ubiquitous in nature
and display a rich and fascinating phenomenology which
remains to be fully elucidated \cite{networks1,networks2}. Coherent macroscopic patterns can
emerge from a limited set of rules which govern the microscopic
evolution of the elementary constituents.
Besides the specificity of the assigned interactions (\textit{e.g.}
long versus short range), global coherence reflects also the
peculiarity of the underlying network of connections. An emblematic, though extreme, example, is the
fully coupled network, where each one of the $N$ nodes experiences a
direct link with all the others. In the opposite limit, the number of
links per node is independent of the size of the network.  Studying
the modifications induced on the statistical properties of the system by such
dilution has important implications for a broad class of applications.

Most studies have so far focused on the finite connectivity scaling,
that is graphs where the number of nodes $N$ and the number of
links $N_L$ go to infinity, keeping the ratio $N/N_L$ fixed.
Models defined on such graphs have been extensively studied in the 
physics literature, especially in the context of spin glasses and combinatorial 
optimization \cite{dorogovtsev,weigt}. Several recent mathematical papers also analyze 
rigorously ferromagnetic Ising models on such graphs~\cite{DeSanctis08,Barra08,BarraContucci08,Agliari08,Dembo08}.

Besides the fully coupled case, comparatively very few papers have been
devoted to the infinite connectivity scaling, where $N,~N_L$ and $N_L/N$
all go to infinity.  Bovier and Gayrard~\cite{bovier} showed
rigorously that, under an appropriate rescaling, the Ising model
defined on such graphs is completely equivalent in the infinite $N$
limit to its fully connected ($N_L=N(N-1)/2$) counterpart. On the basis
of this finding, it can be safely conjectured that analogous
conclusions should apply to other spin models when inspected on a similar
geometry.

The rigorous result by Bovier and Gayrard leaves however open the
question of the speed of convergence towards the fully connected
limit. Finite size effects are indeed crucial when one aims at
understanding the behavior of a finite graph. Given a finite random
graph with relatively high connectivity, two different approaches
may be used to address this issue: on one hand, one might consider
expanding around the fully connected solution of~\cite{bovier}, to
compute the leading order finite $N$ corrections; on the other hand,
the graph under scrutiny can be seen as a realization of an
Erd\"os-R\'enyi random graph \cite{erdos}, with high, but finite, mean
connectivity. This observation enables one to employ the powerful
techniques developed for this latter scaling.

In this paper, we will perform both computations. The first by
resorting to the celebrated replica trick \cite{parisi}; the second, by means of the
cavity method \cite{Par-Mez}. Theoretical estimates will be then compared with
numerical simulations, performed for different sizes and
connectivities, in order to test their accuracy.

The paper is organized as follows. In Section~\ref{section:model},
we introduce the model and present the infinite $N$ solution.  Numerical
simulations are presented in Section~\ref{sec:num}, where the role
of finite size corrections is quantified, with reference to a selected,
macroscopic observable (magnetization).  Section~\ref{replica} is dedicated to
developing the finite size replica calculations, whose predictions are
compared with simulations. The subsequent Section~\ref{cavity}
presents the cavity based analysis.  Finally, in Section~\ref{conclusion} 
we sum up and conclude. Two Appendices, devoted to the annealed model and to the 
Curie-Weiss model, are added in order to give the technical details on 
the textbook developments of Section~\ref{section:model}.

\section{The model}
\label{section:model}

We here consider an Ising model defined on a uniform random network
topology. The network is made of $N$ sites, each tagged
by a discrete counter $i$, which ranges from $1$ to $N$. On each site
sits an Ising spin variable $S_{i}=\pm 1$. Two randomly selected nodes,
say $i$ and $j$, are connected through a coupling constant $J_{ij}$.
The number of links $N_{L}$ is bounded from above by $\widetilde{N}=N(N-1)/2$,
which corresponds to the fully connected case, since we are avoiding
double edging of two sites and self wiring. We introduce the dilution
parameter $\gamma$ by scaling the number of links as
$N_{L}=\binom{N}{\gamma}=N^\gamma/\gamma!\left(1+O(1/N)\right)$, the normalization factor
$\gamma!$ is introduced so that the fully connected topology is exactly 
reproduced when $\gamma \to 2$. We restrict our analysis to the interval $1<\gamma \leq 2$, 
which corresponds to network topologies that range from a number of links growing linearly with
the size of the system ($\gamma=1$) up to the fully connected case
($\gamma=2$).

The Hamiltonian of the Ising model on such a diluted network is
\begin{equation}
\label{hamilt} 
H=-\frac{N}{2N_{L}}\sum_{i\ne j}J_{ij}S_{i}S{j}.
\end{equation}  
With this scaling of the coupling constant the energy is extensive.
In the simplest formulation, $J_{ij}$ is set to an identical reference
value, say $J>0$, if the nodes $i$ and $j$ are connected, zero
otherwise. In the following we shall study the behaviour of the system
as a function of the dilution rate $\gamma$. As proved
in~\cite{bovier}, for $\gamma$ strictly larger than its lower bound
$1$, a second order phase transition of the Curie-Weiss type always occurs. 
Our main goal is to characterize the behaviour of the system for a finite size $N$. 
In the following we start by analyzing the large $N$ limit. The partition
function $Z$ reads
\begin{equation}
\label{part_funct}
Z=\sum_{\lbrace S_{i}\rbrace}\exp \big( {\frac{\beta N}{2N_{L}}\sum_{i\neq j}J_{ij}S_{i}S_{j}} \big)=\\
\sum_{\lbrace S_{i}\rbrace} \prod_{i\neq j}\exp \big({\frac{\beta N}{2N_{L}}J_{i,j}S_{i}S_{j}}\big)~,
\end{equation}
where $\beta$ stands for $1 / k_B T$, where $k_B$ is the Boltzmann
constant and $T$ the system's temperature.  The outer sum in Eq.~(\ref{part_funct}) extends 
over all possible spin configurations. The coupling factor $J_{ij}$ is related to the 
linking probability via the following condition
\begin{equation}
J_{ij}=\Bigg{\{}
\begin{array}{ll}
  J & \text{with probability } p\\
  0 & \text{with probability } 1-p~,\\
\end{array}
\label{probdistrib}
\end{equation}
where $p$ is
\begin{equation}
p=N_{L}/\widetilde{N}= 
  \frac{2}{\gamma!}N^{\gamma-2}\left( 1+O\left( \frac{1}{N}\right)\right)~.
\label{prob}
\end{equation}
The $O(1/N)$ term in the probability represents subdominant
finite size effects, and plays no role in the following. The
probability distribution $P(J_{ij})$ reads
\begin{equation}
\label{probJ}
P(J_{ij})=p\delta(J_{ij}-J)+(1-p)\delta(J_{ij})~.
\end{equation}
We end this section by showing that for all $1<\gamma\leq 2$, this
system is exactly equivalent, in the $N\to \infty$ limit, to the 
fully coupled Curie-Weiss model. This is a non rigorous rephrasing of the main
result in Ref.~\cite{bovier}, which will set the stage for the finite-$N$
studies of the following Sections.

We want to compute $\langle \ln Z\rangle_{J}$, where
$\langle \cdot \rangle_J$ denotes the average over disorder.
This is achieved via the celebrated replica trick, which is based on the identity
\begin{equation}
\langle \ln Z \rangle_{J} = \lim_{n\rightarrow 0}\frac{\langle Z^{n} \rangle_{J}-1}{n}~,
\label{replicatrick}
\end{equation}
where $n$ is a assumed to be a real number. The
central idea consists in carrying out the computation for all
integers $n$, extending the results for all $n$, and performing in
the end the limit for $n\rightarrow 0$.

In our setting the replicated partition function reads
\begin{eqnarray}
Z^{n}&=&\big[\sum_{\{S_{i}\}} \exp \left(-\beta H \right) \big]^{n} \nonumber\\
     &=& \sum_{\{S_{i}^{a}\}} \exp \left( \frac{\gamma!}{2}\beta \frac{1}{N^{\gamma-1}}
     \sum_{a}\sum_{i\neq j}J_{ij}S_{i}^{a}S_{j}^{a} \right) \nonumber \\
     &=&\sum_{\{S_{i}^{a}\}} \exp \left(\frac{\gamma!}{2}\beta\frac{1}{N^{\gamma-1}}
     \sum_{i\neq j}J_{ij}\sum_{a}S_{i}^{a}S_{j}^{a} \right)~,
\end{eqnarray}
where the index $a$ runs over the $n$ replicas.

Averaging over the disorder returns
\begin{eqnarray}
\langle Z^{n}\rangle_{J}&=&\sum_{J_{ij}}P({J_{ij}})\sum_{\{S_{i}^{a}\}} \exp 
        \left(\frac{\gamma!}{2}\beta\frac{1}{N^{\gamma-1}}\sum_{i\neq j}J_{ij}T_{ij} \right) \nonumber \\
	&=& \sum_{\{S_{i}^{a}\}}\sum_{J_{ij}}P({J_{ij}})\prod_{i\neq j} \exp 
	\left(\frac{\gamma!}{2}\beta\frac{1}{N^{\gamma-1}}J_{ij}T_{ij} \right)~,
\end{eqnarray}
where $T_{ij}=\sum_{a}S_{i}^{a}S_{j}^{a}$. Recalling Eq.~(\ref{probJ}), one straightforwardly
obtains
\begin{equation}
\langle Z^{n}\rangle_{J}=\sum_{\{S_{i}^{a}\}}\prod_{i\neq j}\big[1-\frac{2}{\gamma!}
\frac{1}{N^{2-\gamma}}+\frac{2}{\gamma!}\frac{1}{N^{2-\gamma}}e^{\frac{\gamma!}{2}\beta\frac{1}{N^{\gamma-1}}T_{ij}}\big].
\end{equation}
To proceed further, we now expand the exponential function to its
second order approximation, which immediately yields
\begin{eqnarray}
\langle Z^{n}\rangle_{J} &=&\sum_{\{S_{i}^{a}\}}\exp\Big(\sum_{i\neq j}\ln\big[1-\frac{2}{\gamma!}\frac{1}{N^{2-\gamma}} \nonumber \\
                         &+&\frac{2}{\gamma!}\frac{1}{N^{2-\gamma}}\big(1+\frac{\gamma!}{2}\beta\frac{1}{N^{\gamma-1}}T_{ij}
			 +\frac{1}{2}\frac{\gamma!^{2}}{4}\beta^{2}\frac{1}{N^{2\gamma-2}}T_{ij}^{2}+...\big)\Big]\Big),
\end{eqnarray}
and, expanding the logarithm
\begin{equation}
\label{approx}
\langle Z^{n}\rangle_{J}=\sum_{\{S_{i}^{a}\}}\exp\Big(\sum_{i\neq j}\big[\frac{\beta}{N}T_{ij}
+\frac{\gamma!}{4}\frac{\beta^{2}}{N^{\gamma}}T_{ij}^{2}-\frac{1}{2}\frac{\beta^{2}}{N^{2}}T_{ij}^{2}\big]\Big)~.
\end{equation}
Keeping only the leading order in $N$, we have
\begin{eqnarray} 
  \langle Z^{n}\rangle_{J}&=& \sum_{\{S_{i}^{a}\}}\exp\Big(\sum_{i\neq j}\big[ 
\frac{\beta}{N}T_{ij}\big]\Big) \\
&=&\sum_{\{S_{i}^{a}\}}\exp\Big(\frac{\beta}{N}\sum_a \sum_{i\neq j}S_i^aS_j^a 
\Big) \\ 
&=& \left [\sum_{\{S_{i}\}}\exp\Big(\frac{\beta}{N}\sum_{i\neq j}S_iS_j\Big) 
\right]^n \\
&=& \left[ Z_{MF}\right]^n,
\end{eqnarray}
where $Z_{MF}$ is the partition function of the Curie-Weiss
model.  Thus, we recover the fact that at leading order in $N$, the
dilute model is equivalent to the fully coupled one for all
$1<\gamma\leq 2$. The calculation of $Z_{MF}$ is standard and
recalled in Appendix B. We summarize here the results for convenience.
The magnetization of the system $m = \lim_{N \to \infty} \sum_i S_i / N$ 
is obtained by solving the implicit equation
\begin{equation}
m=\tanh (2\beta m)~.
\label{eq:tanh}
\end{equation}
The critical inverse temperature, $\beta_c=1/2$, separates the non
magnetized phase from the magnetized one.  The probability
distribution function of the magnetization can be calculated from
the free energy $F(m)$ as
\begin{equation}
P(m)=\frac{1}{Z(\beta)}\sum_{ \{{\cal C} | m({\cal C})=m \} } \exp \left( -\beta F(m)N \right)~,
\label{eq:Pdim}
\end{equation}
where ${\cal C}$ represents the subsets of spins configurations that have
magnetization equal to $m$. For small~$m$, one can expand $F(m)$ in
powers of $m$ and obtain
\begin{equation}
P(m) \propto \exp \left(-c_2 m^2-c_4 m^{4} \right)~.
\label{eq:Pdim_fin}
\end{equation}
At the critical point $\beta_c=1/2$, one gets $c_2 = 0$ and
$c_4=1/12$.  We stress once again that all these results do not depend
on $\gamma$ in the $N\to \infty$ limit.
 
In the following Section, we will discuss the numerical implementation,
test the above infinite $N$ theory and quantify the finite size
corrections.

\section{Numerical simulations}
\label{sec:num}

The properties of the system are numerically studied
via the Metropolis Monte Carlo algorithm~\cite{montecarlo}. We
focus on the {\it quenched} scenario and reconstruct the average
distribution of the main quantities of interest by averaging over
several realizations of the graph of connections. The analytical 
solution of the {\it annealed} model is sketched in Appendix A. 

\begin{figure}[!h]
\begin{centering}
$\begin{array}{cc}
\includegraphics[width=6.0cm]{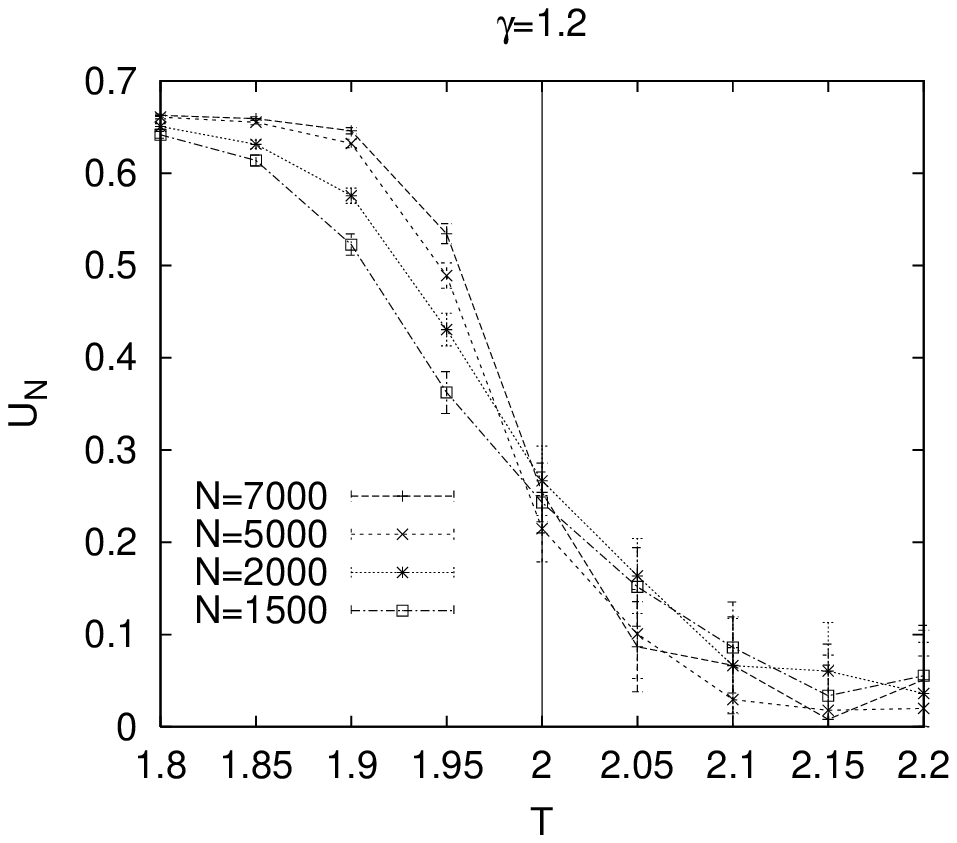}&\includegraphics[width=6.0cm]{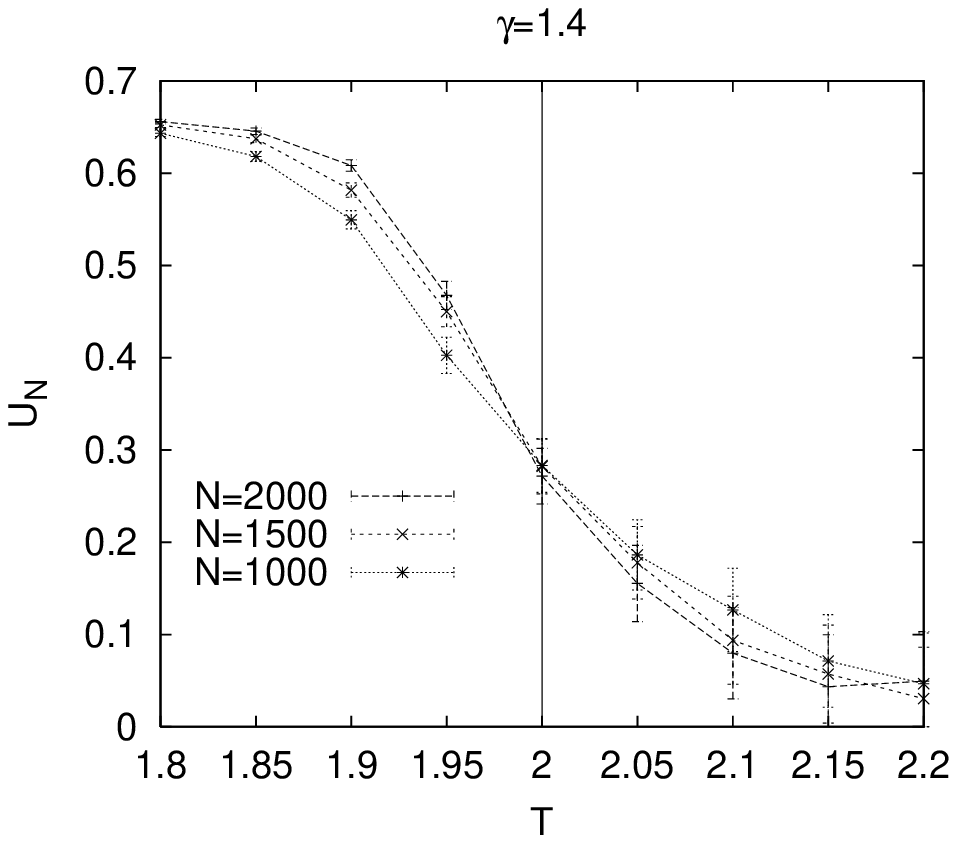}\\
\includegraphics[width=6.0cm]{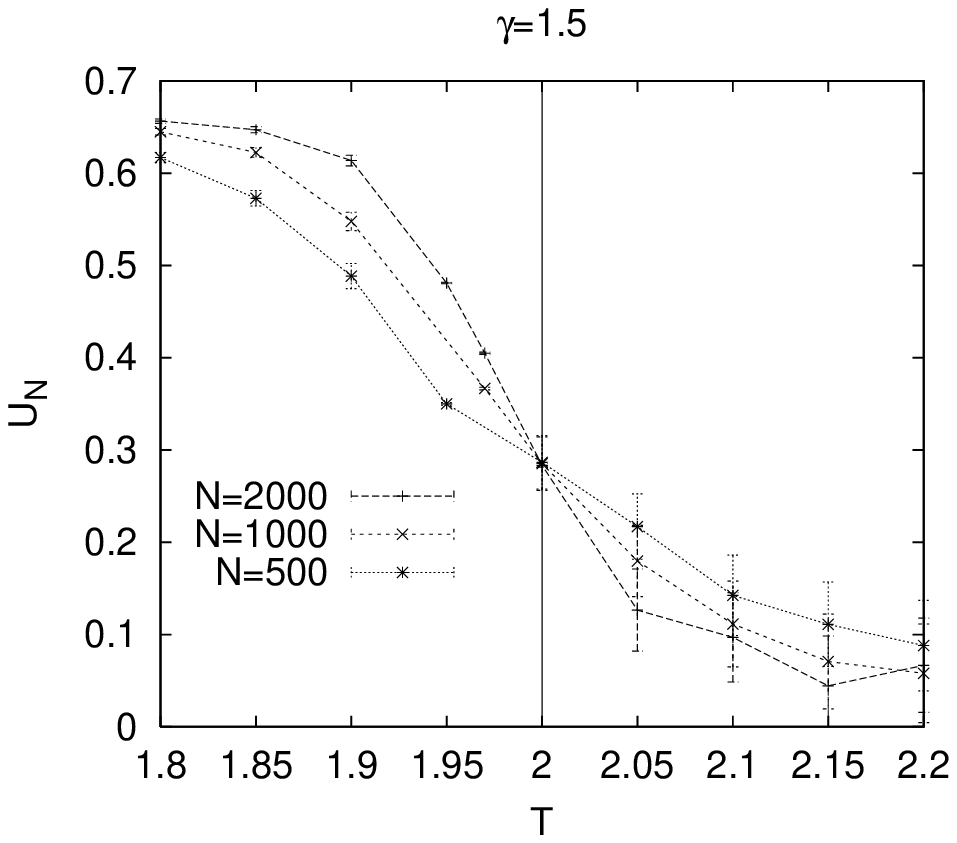}& 
\includegraphics[width=6.0cm]{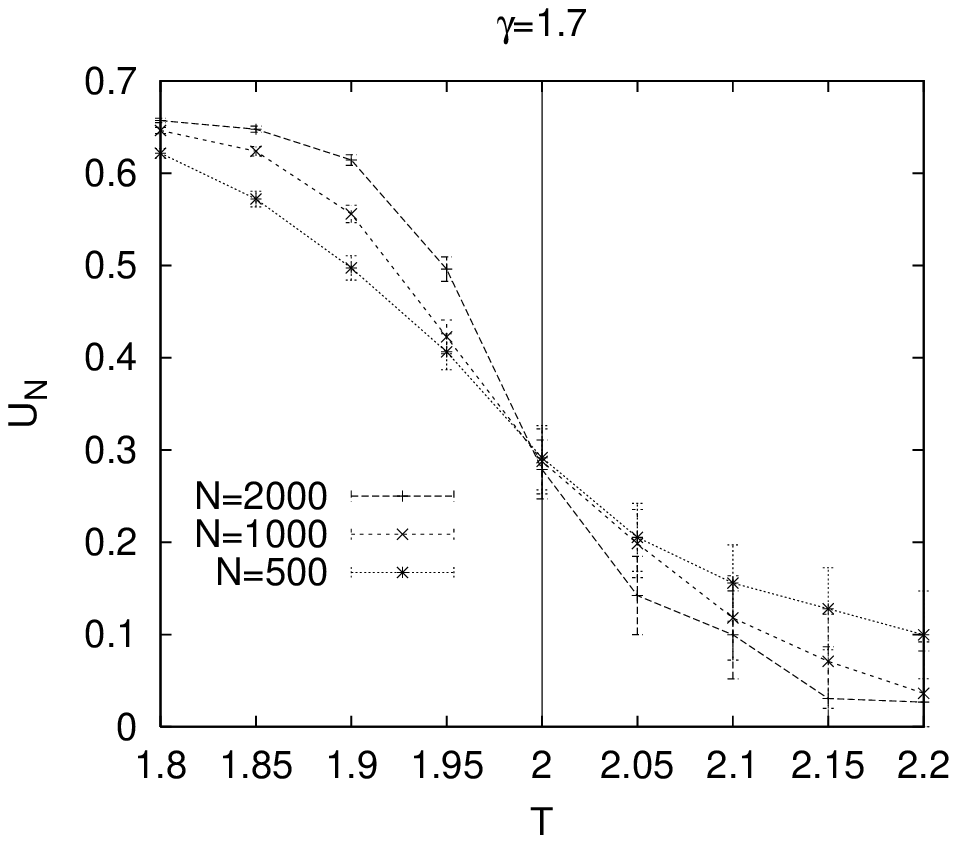}
\end{array}$
\par\end{centering}
\caption{Binder cumulants $U_N$ as a function of temperature $T$, for $\gamma=1.2, 1.4, 1.5, 1.7$. 
The time averages are calculated over $4.10^4$ Monte Carlo sweeps. The error bars are 
estimated with the resampling technique using $10$ sets of $1000$ sweeps and calculating the associated variance.
\label{fig1}}
\end{figure}

The simulated system reproduces well the phase
transition, the actual value of the temperature associated with
symmetry breaking depending on the number of simulated spins.
To test the scenario discussed in the previous Section, we first estimate $\beta_c$
using the so-called Binder cumulant \cite{binder}, defined as
\begin{equation}
U_{N}(T)\equiv1-\frac{\langle m^{4}\rangle}{3\langle m^{2}\rangle^{2}}~,
\end{equation}
where $\langle m^2 \rangle$ and $\langle m^4 \rangle$ denote
respectively the second and fourth moments of the magnetization.  The
Binder cumulant is computed for different values of the imposed
temperature. These numerical experiments are repeated for distinct
values of $N$, while keeping $\gamma$ fixed.  The obtained profiles
$U_N$ vs. $T$ are reported in Fig.~\ref{fig1} for various values of $\gamma$.
Notice that we have introduced a subscript $N$ to recall that the
plotted profiles are reconstructed from finite $N$ calculations. The
importance of the cumulant concept stems from the observation that
curves corresponding to different $N$ all intersect at approximately
the same temperature, which provides an estimate of the critical
temperature $T_c=1/\beta_c$ in the infinite $N$ limit. A direct
inspection of the enclosed figures, suggests that for all values of
$\gamma$ scanning the relevant interval $(1,2)$, $T_c = 2$.  This
result is in agreement with the convergence to the mean-field limit
irrespectively of $\gamma$. Clearly, the convergence to the mean-field
solution is expected to be faster for larger values of $\gamma$.
Indeed, Fig.~\ref{fig1}, panel a), which refers to the case
$\gamma=1.2$, shows a less clear intersection in the interval of $N$
covered by our investigations, when compared to similar plots depicted
for larger values of $\gamma$.

\begin{figure}[t]
   \center
   \includegraphics[width=10cm]{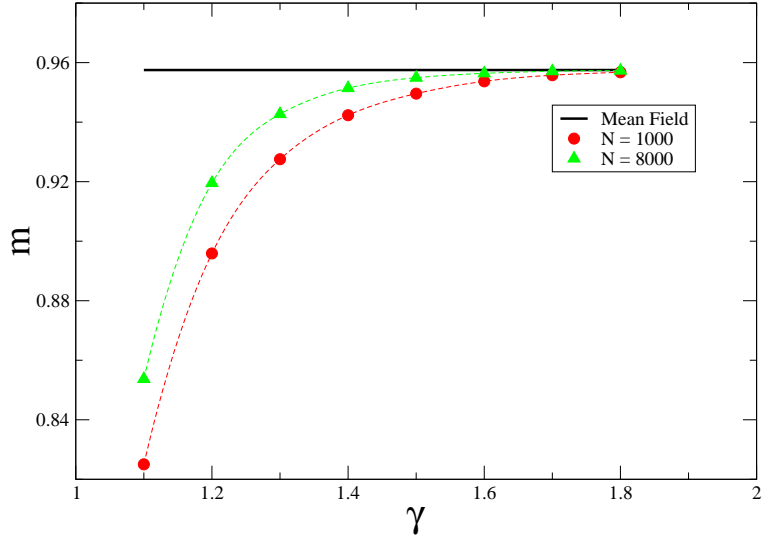}\\
   \caption{Magnetization versus $\gamma$ for $\beta=1$. Symbols refer to the quenched numerical simulations 
   for, respectively, $N=1000$ (circles) and $N=8000$ 
   (triangles). Dashed lines are  guides for the eye.  The solid line stands for the mean-field solution (\ref{eq:tanh}).}
   \label{fig2}
\end{figure}

More interestingly, working at finite $N$, one can monitor the
magnetization and plot it as a function of the
dilution parameter $\gamma$. Results are reported in Fig.~\ref{fig2}
for two different choices of $N$ (symbols): A tendency to
asymptotically approach the mean-field reference (solid) line is
clearly displayed, in agreement with the above scenario. Finite size
corrections play however a crucial role, which deserves to be
carefully addressed. 

Aiming at shedding light onto this issue, we use in the following two
different analytical methods to estimate the finite $N$ corrections.

\section{The replica method}
\label{replica}

The first method relies on the replica trick. Starting with the
calculations of Section~\ref{section:model}, we now include the
leading order finite $N$ corrections. We start again
from Eq.~(\ref{approx}), which we recall here
\begin{equation}
 \langle Z^{n}\rangle_{J}=\sum_{\{S_{i}^{a}\}}\exp\Big(\sum_{i\neq j}
 \big[\frac{\beta}{N}T_{ij}+\frac{\gamma!}{4}\frac{\beta^{2}}{N^{\gamma}}
 T_{ij}^{2}-\frac{1}{2}\frac{\beta^{2}}{N^{2}}T_{ij}^{2}\big]\Big)~.
 \label{approx2}
\end{equation}
The largest neglected term in this expansion is
$\frac{\gamma!^2}{24}\frac{\beta^{3}}{N^{2\gamma-1}}T_{ij}^{3}$;
including it would imply coupling three replicas and make the
calculation technically more difficult. However, this term becomes
progressively more important as $\gamma$ approaches~$1$ from above; 
this may possibly affect the accuracy of the prediction
derived here, in this region of parameters. In the following we shall
also neglect $O(N^{-2})$ terms in expression (\ref{approx2}), which is
certainly well motivated as long as $\gamma<2$ (in a strict sense).

Notice that the following relations apply
 \begin{equation}
   \sum_{i\neq j}T_{ij}=\sum_{i\neq j}\sum_{a}S_{i}^{a}S_{j}^{a}=\sum_{a}
   \sum_{i\neq j}S_{i}^{a}S_{j}^{a}=\sum_{a}\big(\sum_{i}S_{i}^{a}\big)^{2}-Nn=N^{2}\sum_{a}m_{a}^{2}-Nn~,
\label{Tij}
\end{equation}
and
\begin{eqnarray}
\nonumber
\sum_{i\neq j}T_{ij}^{2}&=&\sum_{i\neq j}\sum_{a,b}S_{i}^{a}S_{j}^{a}S_{i}^{b}S_{j}^{b}= \sum_{a,b}\big(\sum_{i,j}S_{i}^{a}S_{j}^{a}S_{i}^{b}S_{j}^{b}-N\big) \\
&=&\sum_{a,b}\big(\sum_{i}S_{i}^{a}S_{i}^{b}\big)^{2}-n^{2}N
=2N^{2}\sum_{a<b}m_{ab}^{2}+nN^{2}-n^{2}N~,
\label{Tij2}
\end{eqnarray}
where use has been made of the definitions
\begin{eqnarray}
m_{a} &=& \sum_i S_{i}^{a} / N~, \\
m_{ab} &=& \sum_{i} S_{i}^{a} S_{i}^{b} / N~.
\end{eqnarray}
Substituting (\ref{Tij}) and (\ref{Tij2}) into Eq.~(\ref{approx}) yields
\begin{equation}
\langle Z^{n}\rangle_{J}= \sum_{\{S_{i}^{a}\}}\exp\Big(\beta N\sum_{a}m_{a}^{2}
-n\beta+\frac{\gamma!\beta^{2}}{2}N^{2-\gamma}\sum_{a<b}m_{ab}^{2}+\frac{\gamma!\beta^{2}}{4}N^{2-\gamma}n\Big)~,
\end{equation}
where the term scaling as $n^{2}$ has been dropped (recall that we
shall be concerned with the limit $n \rightarrow 0$).

The Hubbard-Stratonovich identity can now be invoked to rewrite the
above exponentials involving $m_{a}^{2}$ and $m_{ab}^{2}$
\begin{equation}
\exp \left(\beta N m_{a}^{2}\right)=\sqrt{\frac{\beta N}{\pi}}\int_{-\infty}^{\infty}
d\lambda_{a} \exp \left( -\beta N \lambda_{a}^{2}+2 \beta N m_{a}\lambda_{a} \right)~,
\end{equation}
\begin{eqnarray}
\exp \left(\frac{\gamma!\beta^{2}}{2}N^{2-\gamma}m_{ab}^{2}\right)
&=&\sqrt{\frac{\gamma!\beta^{2}N^{2-\gamma}}{2\pi}}\int_{-\infty}^{\infty}dq_{ab} \nonumber \\ 
&&\exp \left(-\frac{\gamma!\beta^{2}}{2}N^{2-\gamma}q_{ab}^{2}+\gamma!\beta^{2}N^{2-\gamma}m_{ab}q_{ab} \right).
\end{eqnarray}

Putting the various pieces together, the average replicated partition
function reads
\begin{eqnarray}
\label{afterStrat}
\nonumber
&&\langle Z^{n}\rangle_{J}=C\sum_{\{S_{i}^{a}\}}\int_{-\infty}^{\infty}\prod_{a}d\lambda_{a}\\
\nonumber
&&\prod_{a<b}dq_{ab}\exp\Big({-\beta N \sum_{a}\lambda_{a}^{2}+2 \beta N \sum_{a}m_{a}\lambda_{a}
-\frac{\gamma!\beta^{2}}{2}N^{2-\gamma}\sum_{a<b}q_{ab}^{2}+\gamma!\beta^{2}N^{2-\gamma}\sum_{a<b}m_{ab}q_{ab}}\Big) \\
&=&C\int_{-\infty}^{\infty}\prod_{a}d\lambda_{a}\prod_{a<b}dq_{ab}\exp\Big({-\beta N \sum_{a}\lambda_{a}^{2}-\frac{\gamma!\beta^{2}}{2}N^{2-\gamma}\sum_{a<b}q_{ab}^{2}}\Big)\\
\nonumber
&&\sum_{\lbrace S_{i}^{a}\rbrace}\exp\Big({2 \beta N \sum_{a}m_{a}\lambda_{a}+\gamma!\beta^{2}
N^{2-\gamma}\sum_{a<b}m_{ab}q_{ab}}\Big)~,
\end{eqnarray}
where the normalization $C(\gamma,\beta,n,N)=\sqrt{\frac{\beta N}{\pi}}\sqrt{\frac{\gamma!\beta^{2}N^{2-\gamma}}{2\pi}}$ 
can be safely ignored in the forthcoming development.  

Let us focus now on the last sum appearing in Eq.~(\ref{afterStrat}). A straightforward manipulation leads to
\begin{eqnarray}
\nonumber
&&\sum_{\lbrace S_{i}^{a}\rbrace}\exp\Big({2 \beta N \sum_{a}m_{a}\lambda_{a}+\gamma!\beta^{2}N^{2-\gamma}\sum_{a<b}m_{ab}q_{ab}}\Big)\\
&=&\sum_{\{S_{i}^{a}\}}\exp\Big({2 \beta N \sum_{a}\lambda_{a}\sum_{i}\frac{S_{i}^{a}}{N}+\gamma!\beta^{2}N^{2-\gamma}\sum_{a<b}q_{ab}\sum_{i}\frac{S_{i}^{a}S_{i}^{b}}{N}}\Big)\\
&=&\Bigg(\sum_{\{S^{a}\}}\exp\Big({2 \beta 
\sum_{a}\lambda_{a}S^{a}+\gamma!\beta^{2}N^{1-\gamma}\sum_{a<b}q_{ab}S^{a}S^{b}}\Big)\Bigg)^{N}~,
\end{eqnarray}
where the index $i$ can be removed, being replaced by the power $N$.

Let us now introduce the function $\Psi$ as
\begin{equation}
\Psi (\lambda_{a},q_{ab})=\sum_{\{S_{a}\}}\exp\Big({2 \beta N \sum_{a}\lambda_{a}S^{a}
+\gamma!\beta^{2}N^{1-\gamma}\sum_{a<b}q_{ab}S^{a}S^{b}}\Big)\Bigg.~.
\end{equation}
We now make the replica symmetric hypothesis which
corresponds to setting $\lambda_{a}=\lambda$ and $q_{ab}=q$ $\forall
a,b$. After this {\it ansatz}, $\Psi$ can be cast in the form
\begin{equation}
\Psi (\lambda,q)=\sum_{\{S_{a}\}}\exp\Big({2 \beta  \sum_{a}\lambda_{a}S^{a}+\gamma!
\beta^{2}N^{1-\gamma}q\Big[\frac{1}{2}(\sum_{a}S^{a})^{2}-\frac{n}{2}\Big]}\Big).
\end{equation}
The Hubbard-Stratonovich trick allows us to write
\begin{equation}
\exp \left( \frac{\gamma!\beta^{2}}{2}N^{1-\gamma}q(\sum_{a}S^{a})^{2} \right)
=\sqrt{\frac{\gamma!}{2\pi}}\int_{-\infty}^{\infty}dx \exp \left(-\frac{\gamma!}{2}x^{2}+\gamma!\beta 
N^{\frac{1-\gamma}{2}}\sqrt{q}\sum_{a}S^{a}x \right)~,
\end{equation}
which leads to the following expression for $\Psi$
\begin{eqnarray}
\nonumber
\Psi (\lambda,q)&=& \exp \left(-\frac{\gamma!\beta^{2}N^{1-\gamma}qn}{2} \right) \sqrt{\frac{\gamma!}{2\pi}}
\int_{-\infty}^{+\infty} dx \exp \left(-\frac{\gamma!}{2}x^{2} \right) \nonumber \\
&&\sum_{\{S_{a}\}}\exp\Big({(2\beta\lambda+2\gamma!\beta N^{\frac{1-\gamma}{2}}\sqrt{q}x)\sum_{a}S^{a}}\Big) \nonumber \\
&=& \exp \left(-\frac{\gamma!\beta^{2}N^{1-\gamma}qn}{2} \right) \sqrt{\frac{\gamma!}{2\pi}}
\int_{-\infty}^{+\infty}dx \exp \left(-\frac{\gamma!}{2}x^{2} \right) \nonumber \\ 
&& \Big[2\cosh(2\beta\lambda+\gamma!\beta N^{\frac{1-\gamma}{2}}\sqrt{q}x)\Big]^{n}.
\end{eqnarray}
Finally, we obtain
\begin{equation}
\label{Zn}
\langle Z^{n}\rangle_{J}=
C\int_{-\infty}^{\infty}d\lambda dq \exp [N L_{n}]~, 
\end{equation}
where
\begin{equation}
L_n=\Big[-\beta n\lambda^{2}-\frac{\gamma!\beta^{2}}{2}N^{1-\gamma}\frac{n(n-1)}{2}q^{2}+\ln\Psi\Big]~.
\label{eq:Ln}
\end{equation}

\begin{figure}[t]
   \center
   \includegraphics[width=10cm]{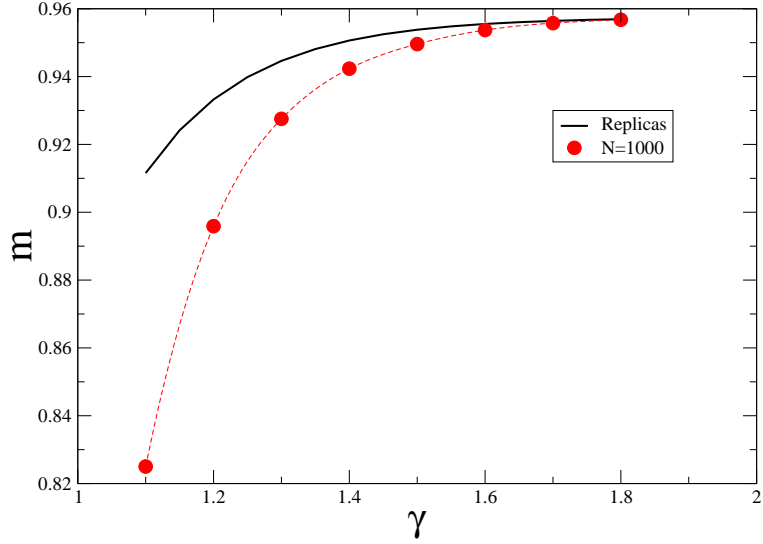}\\
   \caption{Magnetization versus $\gamma$ for $\beta=1$. Symbols refer
     to numerical simulations for $N=1000$ (circles).  The
     dashed lines serves as a guide for the eye.  The solid line
     stands for the replica based solution (\ref{magne_replica}).}
   \label{fig3}
\end{figure}

We want to keep the terms linear in $n$ in Eq.~(\ref{eq:Ln}), since
they are the only ones to give a contribution in the limit $n\to 0$.
A straightforward expansion in $n$ yields, in the $n\to 0$ limit,
\begin{eqnarray}
\frac{1}{n} \ln \Psi \to && -\frac{1}{2}\gamma!\beta^2N^{1-\gamma}q+\ln 2 +
\sqrt{\frac{\gamma!}{2\pi}}\int_{-\infty}^{+\infty} \exp \left(-\gamma!x^2/2 \right) \nonumber \\ 
&& \ln \cosh \left(2\beta \lambda +\gamma!\beta N^{(1-\gamma)/2}\sqrt{q}x\right)~dx~.
\end{eqnarray}
The computation of the free energy is now reduced to finding the saddle 
points of the following function
\begin{eqnarray}
\phi(\lambda,q)&=&-\beta\lambda^2+\frac{\varepsilon^2}{4\gamma!}q^2
-\frac{\varepsilon^2}{2\gamma!}q+\ln 2 +
\sqrt{\frac{\gamma!}{2\pi}}\int_{-\infty}^{+\infty} \exp \left( -\gamma!x^2/2 \right) \nonumber \\
&& \ln \cosh \left(2\beta \lambda +\varepsilon\sqrt{q}x\right)~dx~,
\end{eqnarray}
where we have introduced the small parameter
\begin{equation}
\varepsilon=\gamma!\beta N^{(1-\gamma)/2}~.
\end{equation}
Expanding in $\varepsilon$ up to order $\varepsilon^2$ and performing the Gaussian integrations, we get
\begin{equation}
\phi(\lambda,q)=-\beta\lambda^2+ \ln \cosh 2\beta \lambda + 
\frac{\varepsilon^2}{4\gamma!}q^2
-\frac{\varepsilon^2}{2\gamma!}q \tanh^2 2\beta\lambda +\ln 2
+o(\varepsilon^2)~. 
\end{equation}
At order $\varepsilon^0$, the conditions $\partial_\lambda \phi=0$ and
$\partial_q \phi=0$ yield
\begin{equation}
\lambda_0=\tanh 2\beta\lambda_0~,~q_0=\tanh^2 2\beta \lambda_0~.
\label{implicitLambda}
\end{equation}
As it should, the mean-field solution is recovered from these
equations, see Eq.~(\ref{eq:tanh}). We now write
$\lambda=\lambda_0+\varepsilon^2\lambda_1$, and we get from the
condition $\partial_\lambda \phi=0$
\begin{equation}
\lambda_1=-\frac{1}{\gamma!}\frac{\lambda_0^3(1-\lambda_0^2)}
{1-2\beta(1-\lambda_0^2)}~.
\label{magne_replica}
\end{equation}
The dummy parameter $\lambda$ can be shown to correspond to the
magnetization. Thus, one can compare the replica based prediction
Eq.~(\ref{magne_replica}) (which takes into account the leading order
finite~$N$ corrections) to direct numerical simulations. The comparison
is made in Fig.~\ref{fig3}, where the magnetization is plotted as
a function of $\gamma$. The global trend is captured by
Eq.~(\ref{magne_replica}); however the agreement deteriorates quickly
for small values of $\gamma$. This may be due to the approximations
involved in the calculation.

As a final step, let us compute the leading
finite $N$ correction to the transition temperature. We recall that
its $N\to\infty$ value is $\beta_{c}=1/2$. We compute the Hessian
matrix $H_{\phi}(0,0)$ of $\phi$ in $(\lambda,q)=(0,0)$, getting
\begin{equation}
H_{\phi}(0,0) =
\left(
\begin{array}{cc}
2\beta(2\beta-1) & 0 \\
0 & \frac{\varepsilon^2}{2\gamma!}~.
\end{array}
\right)
\end{equation}
The critical temperature corresponds to a vanishing determinant for
$H_{\phi}(0,0)$. We find, at order $\varepsilon^2$, $\beta_{c}=1/2$.
In conclusion, at this level of approximation, there is no
modification of the critical temperature due to finite $N$ effects.

We end this Section with a comment. A given finite random graph with $N$
sites and $M$ links may be seen as a finite~$N$ realization of a
dilute random graph as above, for some value of $\gamma$. It may also
actually be seen as a finite $N$ realization of a graph constructed
according to the rule described in Eq.~(\ref{prob}) and
Eq.~(\ref{probJ}), with $p=\alpha N_{L}/\widetilde{N}\simeq\frac{2\alpha}{\gamma!}N^{\gamma-2}$.
To each choice of $\gamma \in ]1,2[$ corresponds a value of $\alpha$. There is
then an infinite number of models to which our graph at hand may be
compared. However, the result of a replica calculation at first
subleading order in $1/N$, as performed above for $\alpha=1$, does not
depend on the choice of $\gamma$ and $\alpha$. This freedom thus
cannot be used to optimize our predictions for a finite $N$ graph.

\section{An alternative approach: The cavity method}
\label{cavity}

The method used in the previous Section (expansion in powers of $N$
coupled to a replica calculation) does not give very precise results
for small to moderate values of $\gamma$.  We now turn to an alternative
theoretical approach to interpret the results of our simulations: a
finite size graph with $N$ sites, constructed with the
rule~(\ref{probJ}) for a given $\gamma$, may be seen as a standard
Erd\"os-R\'enyi random graph with parameter
$\lambda_{ER}=N^{(\gamma-1)}/\gamma!$ \cite{erdos,dorogovtsev}. 
This makes possible the use of the powerful methods devised for 
finite connectivity random graphs, such as the cavity method.

The solution of the Ising model on random graphs with arbitrary
distributions of links is given in~\cite{dorog,Leone}; we follow here the
formulation given in~\cite{Par-Mez}, which is there applied to the solution
of an Ising spin glass on a Bethe lattices. Our case is much
simpler, as we are studying a ferromagnet; a small complication is
related to the probability distribution of the site connectivities.

\begin{figure}[t]
   \center
   \includegraphics[width=10cm]{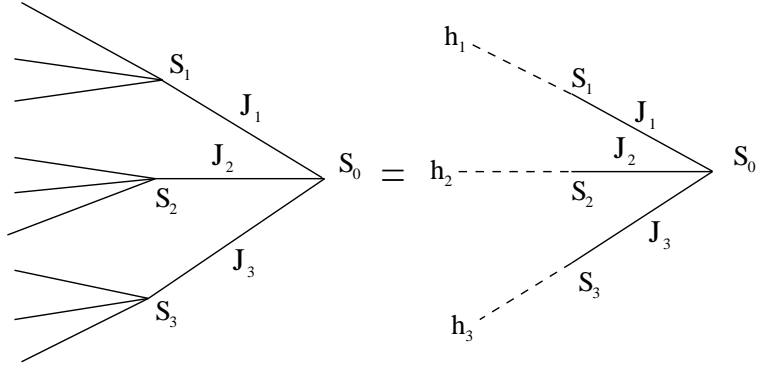}\\
   \caption{This figure shows an example of a tree-like structure with
     $k=3$. The cavity fields $h_{i}$ represent the total field acting
     on the spin $S_{i}$, when the central spin $S_{0}$ is removed.}
   \label{fig_tree}
\end{figure}

We briefly recall the main steps leading to the (replica symmetric)
cavity equations, following Ref.~\cite{Par-Mez}. Consider the Hamiltonian
(\ref{hamilt}) defined on a random graph.  Figure~\ref{fig_tree}
represents a node, denoted with $0$, and its $k=3$ neighbours.  We represent by
$h_i$, $i=1,...,k$ the total field acting on spin $S_i$ {\it in the absence} of
the central spin $S_0$. The magnetization of the $i$-th spin reads
$m_{i}=\tanh(\beta h_{i})$.  The basic ingredient of the cavity method is
to assume that the fields $h_i$ are uncorrelated.

Let us define the partition function of the spin $S_0$ as follows
\begin{equation} 
\label{Z_S_0} 
Z_{S_{0}} = \sum_{S_{0}, S_{1},.., S_{k}}\exp\big(\beta^{'}
S_{0}\sum_{i=1}^{k}J S_{i}+\beta^{'} \sum_{i=1}^{k}h_{i} S_{i}\big)~,
\end{equation}
where
\begin{equation}  
\beta^{'} = \frac{\gamma!}{2}\frac{1}{N^{\gamma-1}}\beta~.
\end{equation}\\
One can now invoke the basic identity
\begin{equation}
  \sum_{S_i=\pm1}\exp\big(\beta^{'}S_0 J S_i+\beta^{'} h_i S_i \big)=c(J,h_i)\exp(\beta^{'} u(J,h_i) S_{0})~, 
\end{equation}
where the two functions $u(\cdot, \cdot)$ and $c(\cdot, \cdot)$ respectively read
\begin{equation}
  u(J,h)=\frac{1}{\beta^{'}}\mbox{arc}\tanh[\tanh(\beta^{'} J)\tanh(\beta^{'} h)]~,
\end{equation}
\begin{equation}
  c(J,h)=2 \frac{\cosh(\beta^{'} J)\cosh(\beta^{'} h)}{\cosh(\beta^{'} u(J,h))},
\end{equation}
and rewrite the partition function (\ref{Z_S_0}) as
\begin{equation} 
\label{Z_S_0_bis} 
Z_{S_0} = \sum_{S_{0},S_{1},.., S_{k}} \Pi_{i=1}^k c(J, k_i) \exp\big(\beta^{'} \sum_{i=1}^{k} u(J,h_i) S_0)~.
\end{equation}

In practice the magnetization on site $0$ is thus given by
$m_{0}=\langle S_{0}\rangle=\tanh(\beta^{'} h_{0})$, where
\begin{equation}
h_{0}=\sum_{i=1}^{k} u(J,h_{i})~.
\label{h_{0}}
\end{equation}
The connectivity $k$ of a given site in the finite size random graph
we study is a random variable with distribution
\begin{equation}
\pi_0(k)=\frac{(N-1)!}{k!(N-1-k)!}(1-N^{\gamma-2})^{N-k}N^{(\gamma-2)k}~.
\label{eq:pdf0}
\end{equation} 
We want to compare our finite size random graph with the corresponding
infinite size graph with the same Erd\"os-R\'enyi parameter
$\lambda_{ER}=N^{\gamma-1}/\gamma!$. This graph has a Poissonian
connectivity distribution with paramter $\lambda_{ER}$. Thus, we use
in the cavity calculations the following distribution, which is close
to~(\ref{eq:pdf0}),
\begin{equation}
\pi(k)=e^{-\lambda_{ER}}\frac{\lambda_{ER}^k}{k\!}~.
\label{eq:pdf}
\end{equation}
Eqs.~(\ref{h_{0}}) and~(\ref{eq:pdf}) allow one to write the following
implicit relation for the probability density $Q(h)$ of local fields
\begin{equation}
Q(h)=\sum_k \pi(k) \int\Pi_{i=1}^{k}[dh_{i}Q(h_{i})]\delta(h-\sum_{i=1}^{k} 
u(J_{i},h_{i})).
\label{distrqu} 
\end{equation}
Eq.~(\ref{distrqu}) can be solved using a population dynamics
algorithm~\cite{Par-Mez} to access an estimate of the local field
distribution $Q(h)$, and eventually compute the magnetization of
the system. More concretely, one starts with arbitrarily chosen
population of $M$ fields and proceeds iteratively as
follows. A random number $k$ is picked up with probability $\pi(k)$; a
subset of $k$ fields is randomly selected in the population, and used
to compute the $h_0$ field, as prescribed by Eq.~(\ref{h_{0}}). Then
one field is removed at random from the population, and replaced with
the computed $h_0$.  Such a scheme defines a Markov chain on the space
of the $M$ fields which admits a stationary distribution. In the limit 
$M \rightarrow \infty$ such a stationary distribution clearly satisfies the self-consistency
relation (\ref{distrqu}). The magnetization $m$ is hence
straightforwardly recovered as $m=\sum_i \tanh(\beta^{'} h_i)/N$.

\begin{figure}[t]
   \center
   \includegraphics[width=10cm]{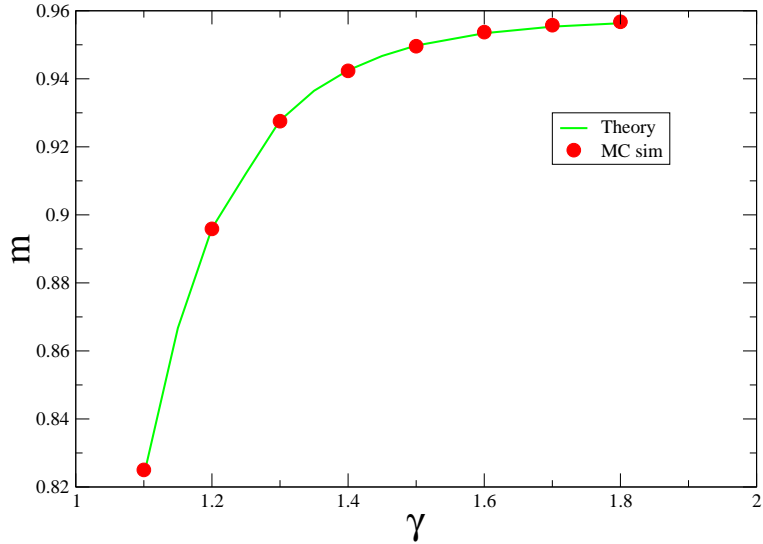}\\
   \caption{Magnetization versus $\gamma$ for $\beta=1$. Symbols refer
     to numerical simulations for $N=1000$ (circles). The
     solid line is the solution obtained using the cavity method. The agreement
     is much better than for the replica approach (see Fig.\ref{fig3}).}
   \label{fig5}
\end{figure}

This method provides very accurate predictions, as shown in
Fig.~\ref{fig5}, much better than the finite $N$ expansion around
the replica calculation discussed in Section~\ref{replica}.

\section{Conclusions}
\label{conclusion}

We have investigated the finite size corrections to
the equilibrium magnetization of the Ising model defined on a diluted network.
Varying the dilution parameter $\gamma$, these networks interpolate between
the fully connected network ($\gamma=2$) and the opposite setting
where the number of links scales linearly with system size $N$
($\gamma=1$).

Systematic deviations with respect to the asymptotic mean field
behavior are observed when a finite number of spins is
considered, such discrepancies being more pronounced as $\gamma$
approaches its lower bound $\gamma=1$.  This phenomenon is clearly displayed
in the plot of the magnetization $m$ versus $\gamma$. A replica based
perturbative analysis is developed, whose predictions are compared with
the outcome of the numerics. The dependence of $m$ on $\gamma$ is
qualitatively captured, but the quantitative match is not satisfying,
especially as the dilution rate is increased. A cavity
based calculation, inspired by the Mezard-Parisi technique \cite{Par-Mez}
is able to reproduce the data with an excellent degree of
accuracy.  

Summing up, we have brought convincing evidences that finite size
corrections do play an important role in presence of a diluted network
and thus need to be carefully addressed. This is best
done by using the cavity metod and the associated population dynamics
algorithm.  Further extension of the present analysis would
include clarifying the reasons why the replica method turns out to be less
accurate than the alternative cavity based approach. This may imply
pushing further the replica calculation by acommodating for the so far
neglected coupling among three independent replicas.

\bigskip
\noindent
{\bf Acknowledgements}

\noindent
We acknowledge financial support from the Galileo project of the Italo-French
University {\it Study and control of models with a large number of interacting
particles}; and from the PRIN07 project {\it Statistical physics of strongly 
correlated systems at and out of equilibrium: exact results and quantum field 
theory methods}.

\newpage

\appendix

\section{Appendix A}
\label{append1}
\begin{figure}[t]
   \center
   \includegraphics[width=10cm]{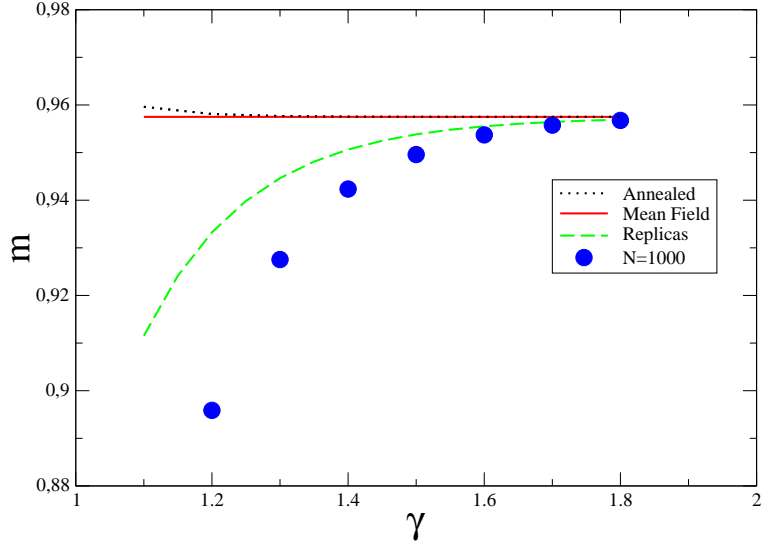}\\
   \caption{Magnetization versus $\gamma$ for $\beta=1$. The solid
     line refers to the mean-field solution; the dashed line stands
     for the replica based calculation; the dotted line represents the
     annealed perturbative estimate.}
   \label{append}
\end{figure}

We shall give here the annealed solution of
model (\ref{hamilt}). Averaging the partition function Eq.~(\ref{part_funct})
over the $J_{ij}$'s, using the probability
distribution~\ref{probdistrib}, one straightforwardly obtains
\begin{align}
  \langle Z\rangle_{J} & =\sum_{\lbrace
    S_{i}\rbrace}\prod_{<i,j>}[1-\frac{2}{\gamma!}N^{\gamma-2}+
  \frac{2}{\gamma!}N^{\gamma-2}e^{\frac{\gamma!\beta}{2N^{\gamma-1}}S_{i}S_{j}}]\nonumber\\
  & =\sum_{\lbrace S_{i}\rbrace}{\exp
    [\sum_{i,j}\ln(1-\frac{2}{\gamma!}N^{\gamma-2}+\frac{2}{\gamma!}N^{\gamma-2}e^{\frac{\gamma!\beta}{2N^{\gamma-1}}S_{i}S_{j}})]}.
  \label{eq:zetamedio}
\end{align}
To proceed further we shall recall that $\gamma>1$, an observation
which in turn allows us to expand in power of $1/N$ the above
expression. The following expression is formally recovered
\begin{equation}
\label{part_funct_expand}
\langle Z\rangle_{J} =\sum_{\lbrace S_{i}\rbrace}\exp \big( {\frac{J \beta_N}{2N}\sum_{i,j}S_{i}S_{j}} \big)~,
\end{equation}
where the finite $N$ temperature, $\beta_N$, reads 
\begin{equation}
\label{beta_N}
 \beta_N = \beta  + J^2 \beta ^3 \left( \frac{\gamma!^2}{24 N^{2 \gamma - 2}} - 
 \frac{\gamma !}{4 N^{\gamma } } + \frac{1}{3 N^{2} }  \right)~.
\end{equation}
In the above derivation we made use of the fact that
$(S_{i}S_{j})^m=1$ for $m$ even and $(S_{i}S_{j})^m=S_{i}S_{j}$
otherwise.  In the limit for $N \rightarrow \infty$, Eq.~(\ref{beta_N}) implies 
$\beta_N \rightarrow \beta$, which in turn implies
\begin{equation}
\label{part_funct_expand1}
\langle Z\rangle_{J} =\sum_{\lbrace S_{i}\rbrace}\exp \big( {\frac{J \beta}{2N}\sum_{i,j}S_{i}S_{j}} \big)~,
\end{equation}
for each value of the $\gamma$ parameter. The
annealed solution is thus also equivalent to the fully coupled
graph solution, in the $N\to\infty$ limit. For finite $N$, Eq.
(\ref{beta_N}) implies a modification of the temperature due to finite
size effects.  According to (\ref{beta_N}), we can imagine to replace
the finite $N$ system at temperature $T=1 / k_B \beta$ with its
Curie-Weiss counterpart, provided a slightly smaller value of the
temperature is allowed. This finding would in turn suggest that the
finite graininess of the distribution drives an increase of the
critical temperature. Consequently, one would expect to observe an
inhomogeneous state at the mean-field transition temperature. This is
at variance with what is found in our (quenched) simulations, where an
opposite tendency is manifested. This discrepancy is also signaled by
inspecting the magnetization as a function of $\gamma$, as outlined in
Fig.~\ref{append}. The numerics in Section \ref{sec:num} and the replica
based analysis of Section \ref{cavity}, that both share the quenched viewpoint, display a similar
trend (though the matching is not perfect as commented in the body of
the paper). Conversely, the annealed prediction obtained from the
mean-field magnetization associated to the finite $N$ temperature
(\ref{beta_N}), returns a striking different behaviour.

\section{Appendix B}
\label{append2} 

Let us here discuss the solution of the mean-field model. By using the
Hubbard-Stratonovich transformation
\begin{equation}
\exp \left( ba^{2} \right)=\sqrt{\frac{b}{\pi}}\int^{+\infty}_{-\infty}dx \exp \left( -bx^{2}+2abx \right)~,
\end{equation}
the partition function (\ref{part_funct_expand1}) (hereafter simply $Z$) can be cast in the form
\begin{eqnarray}
  \qquad\qquad\qquad Z&=&\sum_{\lbrace S_{i}\rbrace} \exp \left( \frac{\beta}{N}(\sum_{i}S_{i})^{2} \right)
  =\sum_{\lbrace S_{i}\rbrace} \exp \left( \frac{N^{2}\beta}{N}(\frac{(\sum_{i}S_{i})}{N})^2 \right)
  =\nonumber\\
  &=&\sqrt{\frac{\beta N}{\pi}}\sum_{\lbrace S_{i}\rbrace}\int^{+\infty}_{-\infty}dx \exp \left( 
  -\beta Nx^{2}+2\beta N \sum_{i}\frac{S_{i} x}{N} \right)=\nonumber\\
  &=&\sqrt{\frac{\beta N}{\pi}}\int^{+\infty}_{-\infty}dx \exp \left(-\beta Nx^{2} \right)
  \sum_{\lbrace S_{i}\rbrace} \exp \left( 2\beta N \sum_{i}\frac{S_{i} x}{N} \right)~,
\label{eq:Z(x)}
\end{eqnarray}
where $a=(\sum_{i}S_{i})^{2}$ and $b=\beta N$ and eventually 
\begin{equation}
 Z=\sqrt{\frac{\beta N}{\pi}}\int^{+\infty}_{-\infty}dx \exp \left( -N(\beta x^{2}-\ln(2\cosh(2\beta x))) \right)~.
\end{equation}
The free energy function results in
\begin{equation}
  -\beta F= \lim_{N\rightarrow\infty}\frac{1}{N}\ln Z_{N}=\lim_{N\rightarrow\infty}\frac{1}{N}\ln
  \Big[ \sqrt{\frac{\beta N}{\pi}}\int^{+\infty}_{-\infty}dx \exp \left( -N(\beta x^{2}-\ln(2\cosh(2\beta x))) \right) \Big]~,
\end{equation}
and using the saddle point approximation
\begin{equation}
-\beta F=\max_{x}[-\beta x^{2}+\ln(2\cosh(2\beta x))].
\label{eq:betaF}
\end{equation}

\end{document}